\input{aipcheck}

\documentclass[
  ]
  {aipproc}

\layoutstyle{8x11single}

\begin{document}

\title{TMD PDFs in Drell-Yan lepton pair production at LHC}


\author{S.P. Baranov}{
  address={Lebedev Institute of Physics, 119991 Moscow, Russia}
}

\author{A.V. Lipatov}{
  address={SINP, Lomonosov Moscow State University, 119991 Moscow, Russia \\
Joint Institute of Nuclear Physics, 141980 Dubna, Moscow Region, Russia}
}

\author{\underline{N.P. Zotov}}{
  address={SINP, Lomonosov Moscow State University, 119991 Moscow, Russia}
%%  ,altaddress={<author1 address>} % additional visiting address
}

\begin{abstract}
We consider the transverse momentum dependent (TMD) quark densities of the
proton which are very important ingredients for unpolarized Drell-Yan (DY) lepton pair production. We
calculate the TMD sea quark density as a convolution of the
Catani-Ciafaloni-Fiorani-Marchesini (CCFM)-evolved gluon distribution and the TMD gluon-to-quark splitting function.
 Based on the ${\cal O}(\alpha^2)$ production amplitude
$q^* + \bar q^* \to Z/\gamma^* \to l^+ + l^-$, calculated by taking into account the effective $q^*\bar q^* Z/\gamma^*$ - vertex, we analyze the distributions on the dilepton invariant mass, transverse momentum and rapidity and specific angular correlations between the produced leptons as measured by the CMS, ATLAS and LHCb collaborations.
   We briefly duscuss also the process of assiciated lepton pair and jet production in $pp$ collisions at the LHC.
\end{abstract}
%\end{document}
%\endinput
\maketitle

%%%%%%%%%%%%%%%%%%%%%%%%%%%%%%%%%%%%%%%%%%%%
%% MAINMATTER
%%%%%%%%%%%%%%%%%%%%%%%%%%%%%%%%%%%%%%%%%%%%

\section{Introduction}
The production of Drell-Yan lepton pairs at the LHC
is subject of intense studies from both
theoretical and experimental points of view.
It provides a major source of background to a number of processes, such as Higgs, $t\bar t$-pair, di-boson or $W^\prime$ and $Z^\prime$ bosons production and other processes beyond the Standard Model.

  The pragmatic goal of our research is to obtain within
the framework of the $k_T-$factorization approach the 
description of measured cross sections close to the precision one, in particular in the forward region, where the 
small-$x$ dynamics come to the game. The predictive power of 
the $k_T-$factorization approach largely depends on the transverse momentum depenent (TMD) parton distributions~\cite{Col}. So far the CCFM TMD parton distributions in the framework of the $k_T$-factorization approach take only gluon and valence quark contributions into account. However, the description of the quark-initiated processes requires to use the TMD sea quark densities. Recently, the TMD sea quark densities have been calculated~\cite{HHJ} incorporating the effects of the TMD gluon-to-quark splitting
function~\cite{CH} which contains all single logarithmic small-$x$ corrections to sea quark evolution for any order of perturbation theory. In the present paper we apply the TMD sea quark densities~\cite{HHJ} to investigate the Drell-Yan lepton pair production and their associative one with jet at the LHC. Unlike the papers~\cite{WMR,SS,LMZ} we concentrate on the off-shell quark-antiquark annihilation $q^* + \bar q^* \to Z/\gamma^* \to l^+ + l^-$ and calculate the corresponding production amplitude according to the reggeized quark approach~\cite{LV,BF}, which is based on the effective
action formalism~\cite{L} in addition to accounting of transverse momentum dependence in the gluon-to-quark branching~\cite{CH}.   
 
\section{Theoretical framework}
 Our consideration is based on the ${\cal{O}}(\alpha^2)$ subprocess of off-shell quark-antiquark annihilation into a virtual photon or $Z$ boson  which decays to lepton pair:
\begin{equation}
  q^*(q_1) + \bar q^*(q_2) \to Z/\gamma^* \to l^+(p_1) + l^-(p_2),
\end{equation}

\noindent
where the four-momenta of all corresponding particles are given in the parentheses. In our theoretical calculations we follow the paper~\cite{BLZ}.

 To calculate the total and differential cross sections
one has to convolute the evaluated off-shell amplitude squared $|\bar {\cal M}|^2$ with the
TMD quark densities of the proton. Our master formula reads:
\begin{equation}
  \sigma=\sum_q\int\frac{|\bar {\mathcal M}|^2}{16\pi\, (x_1 x_2 s)^2} f_{q}(x_1
,\mathbf q_{1T}^2,\mu^2) f_{q}(x_2,\mathbf q_{2T}^2,\mu^2) d\mathbf p_{1T}^2 d\mathbf q_{1T}^2 d\mathbf q_{2T}^2 dy_1 dy_2 \frac{d\phi_1}{2\pi}\frac{d\phi_2}{2\pi},
\end{equation}

\noindent
where $s$ is the total energy, $y_1$ and $y_2$ are the center-of-mass rapidities of the produced leptons, $\phi_1$ and
$\phi_2$ are the azimuthal angles of the initial quarks having the fractions $x_1$ and $x_2$
of the longitudinal momenta of the colliding protons. The 
off-shell matrix elements were done in~\cite{BLZ}.
\begin{figure}[!b]
  \includegraphics[height=.37\textheight,angle=270]{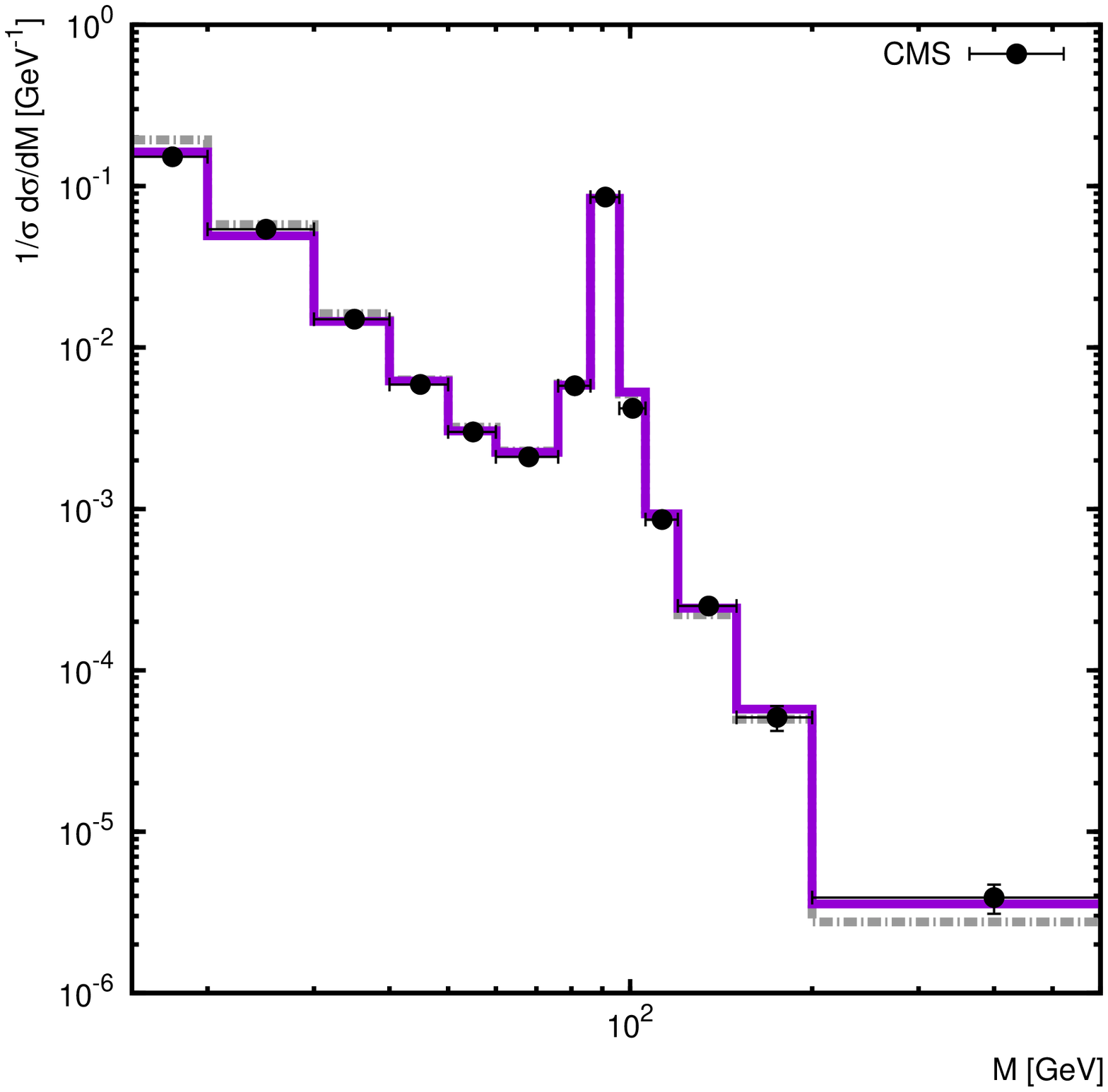}
%  \hspace*{.1cm}
  \includegraphics[height=.37\textheight,angle=270]{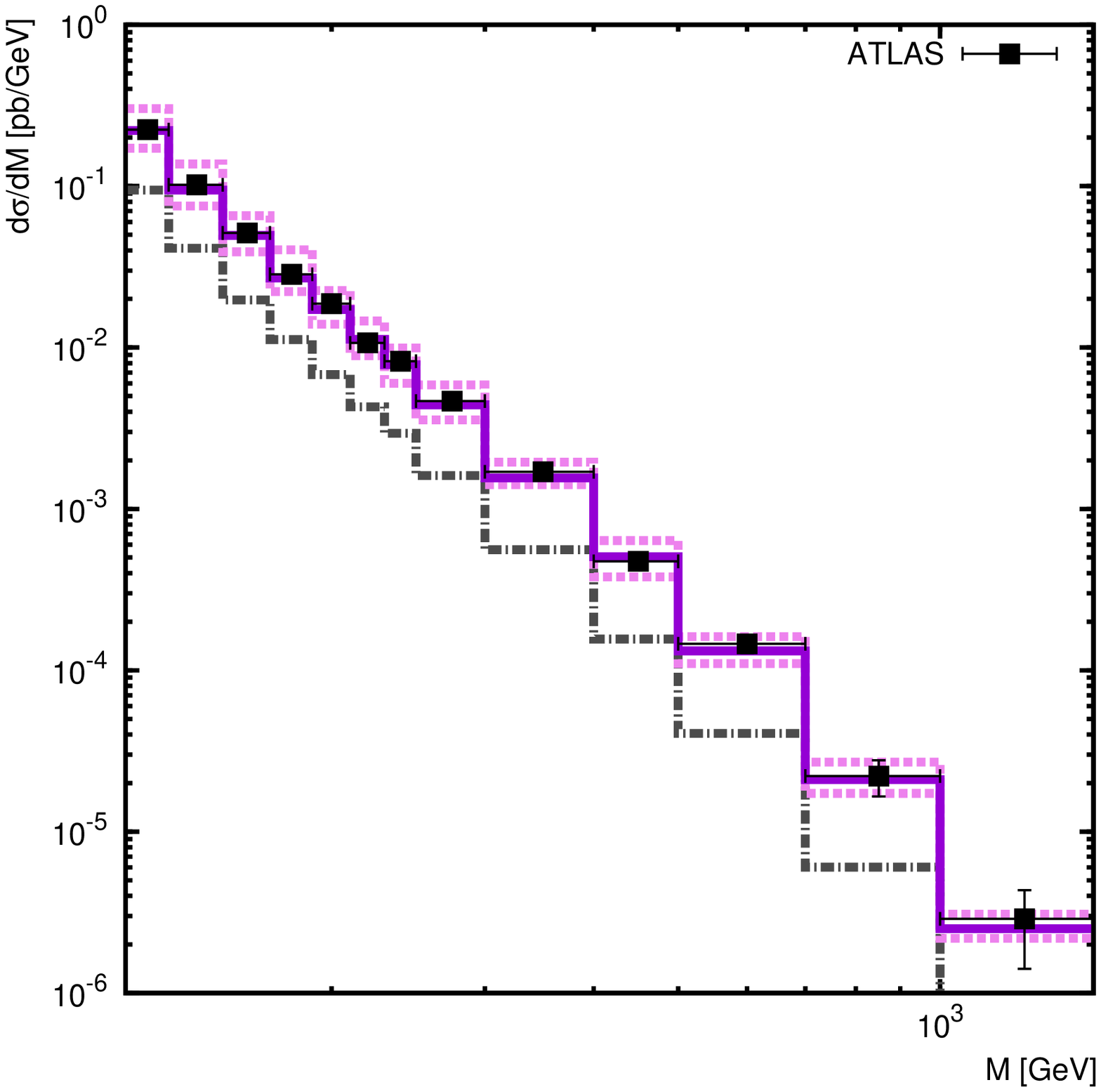}
  \caption{The differential cross sections of Drell-Yan lepton pair production in $pp$ collisions at the LHC as a function of dilepton invariant mass. The solid and dash-dotted histograms correspond to the CCFM-based and KMR predictions,
respectively. The upper and lower dashed histograms correspond to the scale variations in the CCFM calculations, as it is described in the text.
The experimental data are from CMS~\cite{25} and ATLAS~\cite{31}.}
\label{fig1}
\end{figure}
Here we concentrate on the CCFM
approach to calculate the TMD parton densities of the proton.
The TMD gluon~\cite{Jung} and valence quark~\cite{DJK}
distributions $f_g(x,{\mathbf k}_T^2,\mu^2)$ and
$f_q^{(v)}(x,{\mathbf q}_T^2,\mu^2)$ have been obtained
from the numerical solutions of the CCFM equation.
Here ${\mathbf k}_T$ and ${\mathbf q}_T$ are the
gluon and quark transverse momenta, respectively.
In the approximation where the sea quarks occur in the last gluon-to-quark splitting,
the TMD sea quark density at the next-to-leading logarithmic accuracy $\alpha_s(\alpha_s \ln x)^n$ can be written\cite{HHJ} as follows:
\begin{equation}
  f_q^{(s)}(x,{\mathbf q}_T^2,\mu^2) = \int \limits_x^1 {dz \over z} \int d{\mathbf k}_T^2
    {1\over {\mathbf Q}^2} {\alpha_s \over 2\pi} P_{qg}(z,{\mathbf k}_T^2,{
\mathbf Q}^2) f_g(x/z,{\mathbf k}_T^2, \bar \mu^2),
\end{equation}

\noindent
where $z$ is the fraction of the gluon light cone momentum which is carried out by the quark, and ${\mathbf Q} = {\mathbf q}_T - z{\mathbf k}_T$.
The sea quark evolution is driven by the off-shell  gluon-to-quark splitting
function $P_{qg}(z,{\mathbf k}_T^2,{\mathbf Q}^2)$~\cite{CH}:
\begin{equation}
  P_{qg}(z,{\mathbf k}_T^2,{\mathbf Q}^2) = T_R \left({\mathbf Q}^2\over {\mathbf Q}^2 + z(1-z)\,{\mathbf k}_T^2\right)^2
    \left[(1 - z)^2 + z^2 + 4z^2(1 - z)^2 {{\mathbf k}_T^2\over {\mathbf Q}^2} \right],
\end{equation}

\noindent      
with $T_R = 1/2$. The splitting function $P_{qg}(z,{\mathbf k}_T^2,{\mathbf Q}^2)$
has been obtained by generalizing to finite
transverse momenta, in the high-energy region, the two-particle irreducible kernel expansion~\cite{CFP}.
The scale $\bar \mu^2$ is defined\cite{Jung}
from the angular ordering condition which is natural from the point of view of the CCFM evolution: $\bar \mu^2 = {\mathbf Q}^2/(1-z)^2 + {\mathbf k}_T^2/(1-z)$.
To be precise, in~(3) we have used A0 gluon~\cite{Jung}.
\begin{figure}[!b]
  \includegraphics[height=.37\textheight,angle=270]{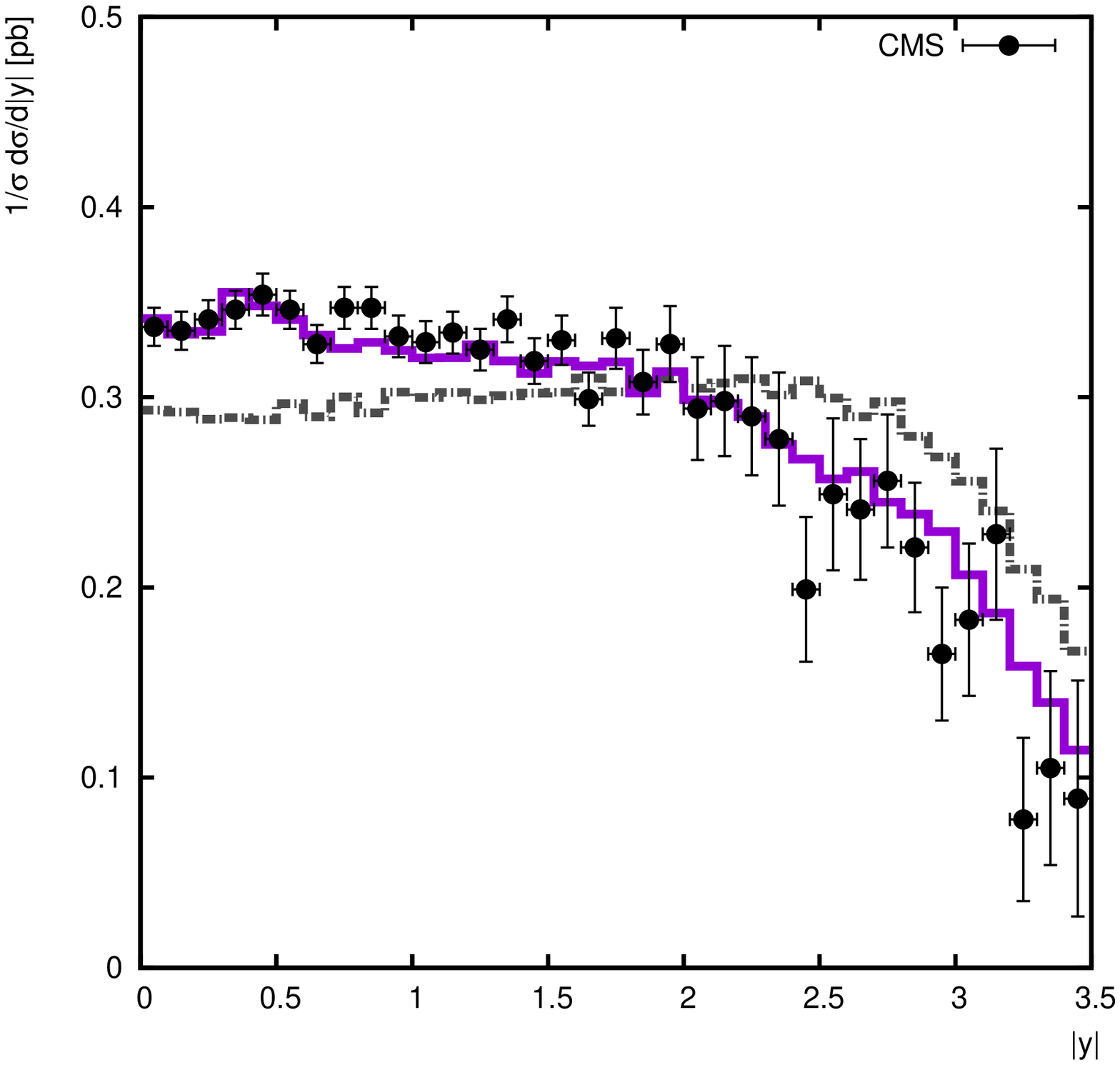}
%  \hspace*{.1cm}
  \includegraphics[height=.37\textheight,angle=270]{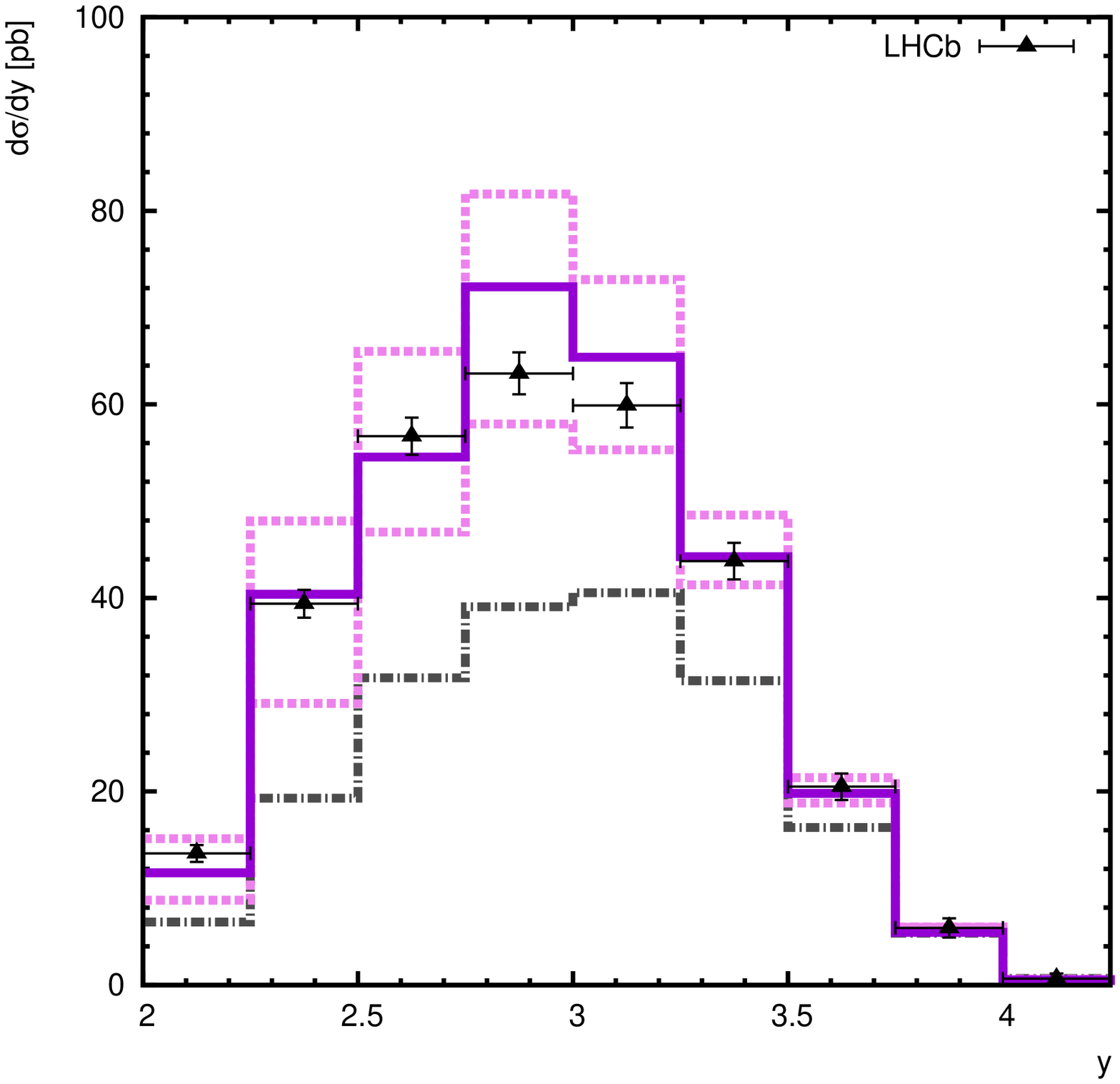}
  \caption{The differential cross sections of Drell-Yan lepton pair production in $pp$ collisions at the LHC as a function of dilepton rapidity $y$. Notation of all histograms is the same as in Fig.~1. The experimental data are from CMS~\cite{26} and LHCb~\cite{32}.}
\label{fig2}
\end{figure}

Beside the CCFM-based approximation above,
to determine the TMD quark densities in a proton
we have used also the Kimber-Martin-Ryskin (KMR) approach\cite{KMR,WMR}. The difference between the CCFM-based and KMR approaches are visible  clearly (see~\cite{BLZ}) in the sea quark distributions which are driven mainly by the gluon densities.

\section{Numerical results}
\begin{figure}[!b]
  \includegraphics[height=.37\textheight,angle=270]{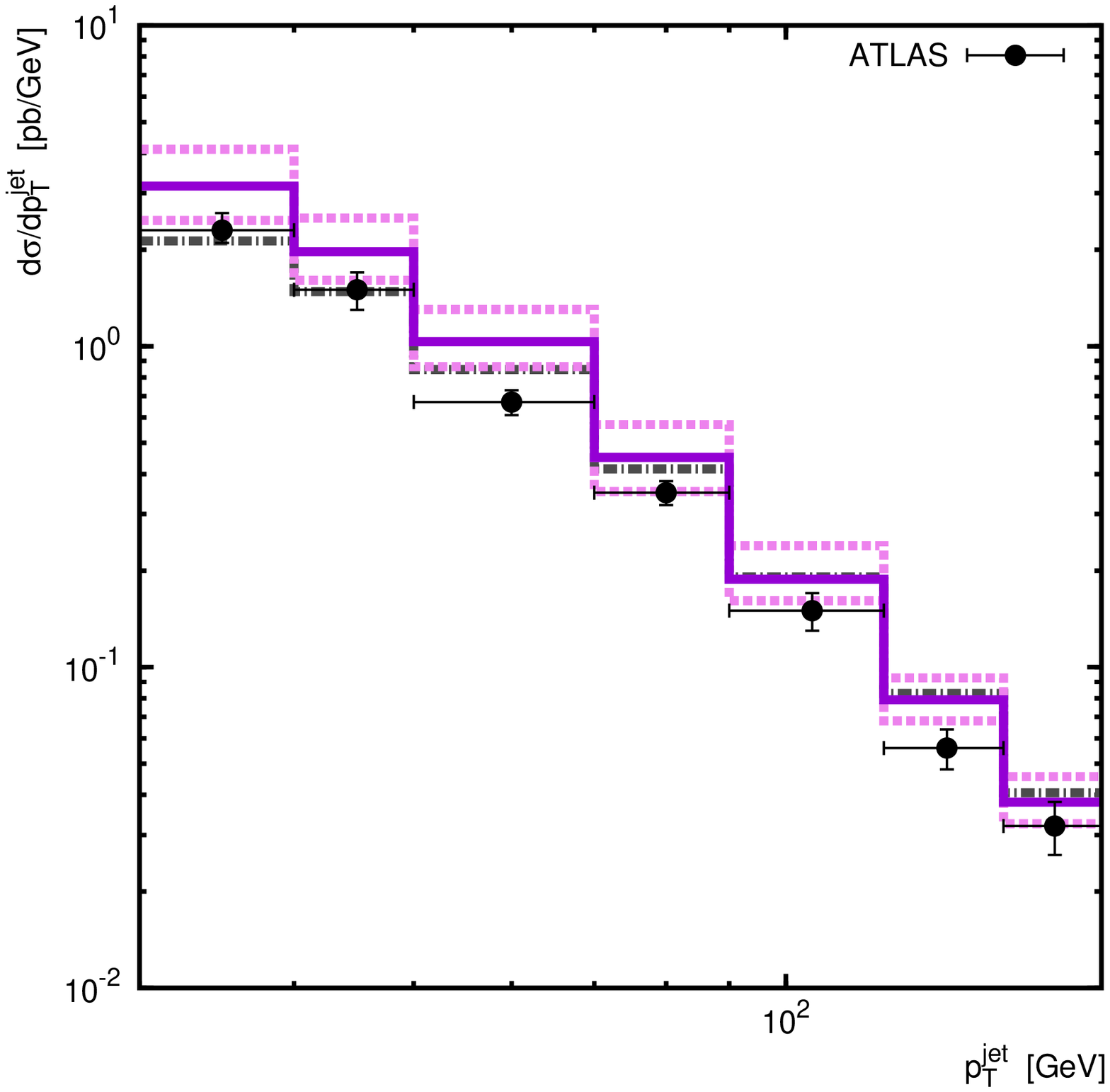}
%  \hspace*{.1cm}
  \includegraphics[height=.37\textheight,angle=270]{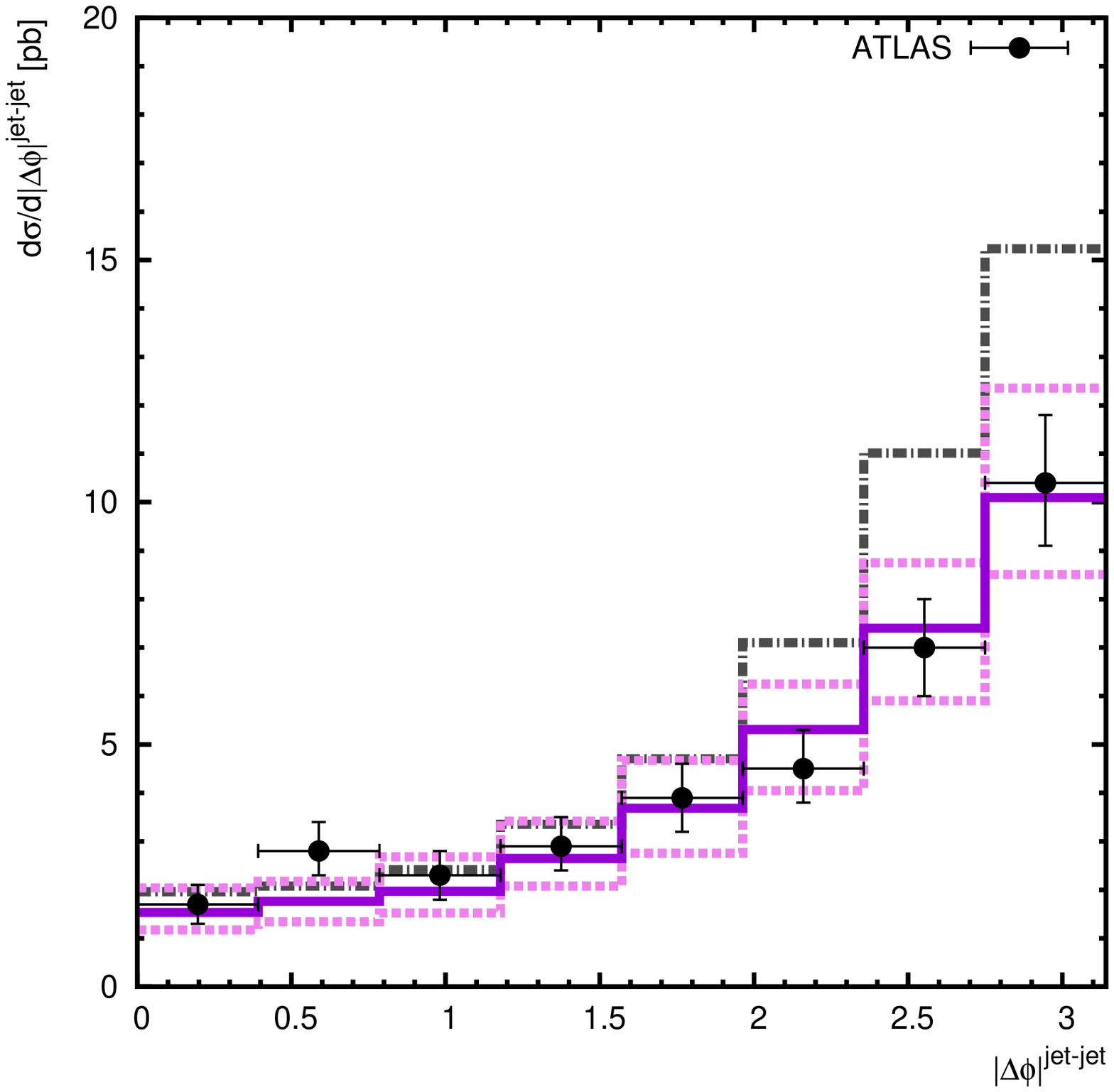}
  \caption{The differential cross sections of associated lepton pair production in $pp$ collisions at the LHC. Notation of histograms is the same as in Fig. 1.
The experimental data are from ATLAS \cite{29}.}
\end{figure}
After we fixed the TMD quark densities, the cross section~(2) depends on the renormalization and factorization scales $\mu_R$ and $\mu_F$. Numerically, we set them to be equal to $\mu_R = \mu_F = \xi M$, where $M$ is the invariant mass of produced lepton pair.
To estimate the scale uncertainties of our calculations we vary the parameter
$\xi$ between $1/2$ and $2$ about the default value $\xi = 1$.
 We set $m_Z = 91.1876$~GeV, $\Gamma_Z = 2.4952$~GeV,
$\sin^2 \theta_W = 0.23122$ and use the LO formula for the strong coupling constant
$\alpha_s(\mu^2)$ with $n_f= 4$ active quark flavors at
$\Lambda_{\rm QCD} = 200$~MeV, so that $\alpha_s (m_Z^2) = 0.1232$.
Since we investigate a wide region of $M$, we use the running QED coupling
constant $\alpha(\mu^2)$. To take into account the non-logarithmic loop corrections to the
quark-antiquark annihilation cross section we apply the effective $K$-factor, as it was done in~\cite{WMR,KS}:
\begin{equation}
  K = \exp \left[ C_F {\alpha_s(\mu^2)\over 2\pi} \pi^2 \right],
\end{equation}

\noindent
where color factor $C_F = 4/3$. A particular scale choice
$\mu^2 = {\mathbf p}_T^{4/3} M^{2/3}$
(with ${\mathbf p}_T$ being the transverse momentum of
produced lepton pair) has been proposed~\cite{WMR,KS}.

Some of the results of our calculations are presented in Figs. 1 --- 2 in comparison with the LHC data (for more details see~\cite{BLZ}).
The differential cross sections as a function of dilepton invariant mass and rapidity are shown in Figs.~2 and~3. We find that these distributions are
described reasonably well by the CCFM-based calculations.
However, the KMR predictions significantly (by a factor of about 2) underestimate the LHC data,
mainly due to different behaviour of corresponding TMD sea quark densities at low transverse
momenta (see Fig.~1). We observe that the shape of dilepton invariant mass distributions is not very sensitive to the TMD quark densities. This is in a contrast with the distributions on the dilepton rapidity, where the CCFM and KMR predictions differ from each other  and the KMR TMD does not describe 
the data (Fig.~2).

Fig. 3 shows our results~\cite{LZ} for the associated llepton pair and jet production in $pp$ collisions at the LHC. We see that the CCFM-based predictions agree well with the data $\Delta^{jet-jet}$ distribution, where the role
of jets originated from the evolution cascade, which results in the uncertainties
in our predictions (see discuss in~\cite{LZ}), is suppressed due to hard scale $\mu^2 \sim m_Z^2$.

To conclude, we have demonstrated that the studies of Drell-Yan lepton pair production and also in the association with jet in $pp$ collisions at the LHC provide important information about the TMD quark distributions of the proton.
The sensivity of predicted cross sections to the TMD quark densities is clearly visible in the mass and the rapidity distributions of the produced lepton pairs 
and the jet-jet correlations. It is important for futher investigations of
 small-$x$ physics at hadron colliders, in particular, in the direction which concerns the non-linear effects originating from high parton densities
 at small $x$.

%%%%%%%%%%%%%%%%%%%%%%%%%%%%%%%%%%%%%%%%%%%%%%%%
%% BACKMATTER
%%%%%%%%%%%%%%%%%%%%%%%%%%%%%%%%%%%%%%%%%%%%%%%%

\begin{theacknowledgments}
We thank H.~Jung for careful
reading the manuscript and very useful remarks.
The authors are also grateful to F.~Hautmann and S.~Marzani for
discussions and comments.
This research was supported by the FASI of Russian Federation
(grant NS-3042.2014.2), RFBR grant 13-02-01060 and the grant of the
Ministry of education and sciences of Russia (agreement 8412).
The authors are also grateful to DESY Directorate for the
support in the framework of Moscow---DESY project on Monte-Carlo implementation
for HERA---LHC. N.Z. is very grateful to the Organization Committee, in particular A. Papa and R. Fiore,  for the invitation and the financial support.  
\end{theacknowledgments}



\bibliographystyle{aipproc}   % if natbib is available

\bibliography{sample}

\IfFileExists{\jobname.bbl}{}
 {\typeout{}
  \typeout{******************************************}
  \typeout{** Please run "bibtex \jobname" to optain}
  \typeout{** the bibliography and then re-run LaTeX}
  \typeout{** twice to fix the references!}
  \typeout{******************************************}
  \typeout{}
 }

%\end{document}

%%%%%%%%%%%%%%%%%%%%%%%%%%%%%%%%%%%%%%%%%%%
%% The following lines show an example how to produce a bibliography
%% without the help of the BibTeX program. This could be used instead
%% of the above.
%%%%%%%%%%%%%%%%%%%%%%%%%%%%%%%%%%%%%%%%%%%

\end{document}